\documentclass[prl,aps,twocolumn,superscriptaddress,showpacs,floatfix,tightenlines]{revtex4}

\usepackage{amsmath}
\usepackage{amsfonts}
\usepackage{epsfig}
\usepackage[dvips]{color}

\begin{document}

\title{Quantum process tomography of a controlled-{\sc not} gate}

\author{J.~L.~O'Brien}
\homepage{www.quantinfo.org}
\affiliation{Centre for Quantum Computer Technology and Physics Department, University of Queensland, Brisbane 4072, Australia}
\author{G.~J.~Pryde} 
\homepage{www.quantinfo.org}
\affiliation{Centre for Quantum Computer Technology and Physics Department, University of Queensland, Brisbane 4072, Australia}
\author{A.~Gilchrist}
\affiliation{Centre for Quantum Computer Technology and Physics Department, University of Queensland, Brisbane 4072, Australia}
\author{D.~F.~V.~James}
\affiliation{Theoretical Division T-4, Los Alamos National Laboratory, Los Alamos, NM 87545, USA.}
\author{N.~K.~Langford}
\homepage{www.quantinfo.org}
\affiliation{Centre for Quantum Computer Technology and Physics Department, University of Queensland, Brisbane 4072, Australia}
\author{T.~C.~Ralph}
\affiliation{Centre for Quantum Computer Technology and Physics Department, University of Queensland, Brisbane 4072, Australia}
\author{A.~G.~White}
\affiliation{Centre for Quantum Computer Technology and Physics Department, University of Queensland, Brisbane 4072, Australia}

\begin{abstract}
We demonstrate complete characterization of a two-qubit entangling process -- a linear optics controlled-{\sc not} gate operating with coincident detection -- by quantum process tomography. We use a maximum-likelihood estimation to convert the experimental data into a physical process matrix.  The process matrix allows accurate prediction of the operation of the gate for arbitrary input states and calculation of gate performance measures such as the average gate fidelity, average purity and entangling capability of our gate, which are 0.90, 0.83 and 0.73 respectively. 
\end{abstract}
\maketitle

Quantum information science offers the potential for major advances such as quantum computing \cite{mikenike} and quantum communication \cite{gisin}, as well as many other quantum technologies \cite{dowlingmilburn}. Two-qubit entangling gates, such as the controlled-{\sc not (cnot)}, are fundamental elements in the archetypal quantum computer \cite{mikenike}. A promising proposal for achieving scalable quantum computing is that of Knill, Laflamme and Milburn (KLM), in which linear optics and a measurement-induced Kerr-like nonlinearity can be used to construct {\sc cnot} gates \cite{KLM}. Gates such as these can also be used to prepare the required entangled resource for optical cluster state quantum computation~\cite{cluster}. The nonlinearity upon which the KLM and related \cite{HiFi,imoto} {\sc cnot} schemes are built can be used for other important quantum information tasks, such as quantum nondemolition measurements \cite{QTqnd,kokQND} and preparation of novel quantum states (for example, \cite{QTmultiport}). An essential step in realizing such advances is the complete characterization of quantum processes. 

A complete characterization in a particular input/output state space requires determination of the mapping from one to the other. In discrete-variable quantum information, this map can be represented as a {\em state} transfer function, expressed in terms of a {\em process matrix} $\chi$. Experimentally, $\chi$ is obtained by performing {\em quantum process tomography} (QPT) \cite{chuangQPT,poyatos}. QPT has been performed in a limited number of systems. A one-qubit teleportation circuit \cite{nielsenNMR}, and a controlled-NOT process acting on a highly mixed two-qubit state \cite{childs} have been investigated in liquid-state NMR. In optical systems, where pure qubit states are readily prepared, one-qubit processes have been investigated by both ancilla-assisted \cite{altepeter,deMartini} and standard \cite{nambu} QPT. Two-qubit optical QPT has been performed on a beamsplitter acting as a Bell-state filter \cite{mitchell}. 

We fully characterize a two-qubit entangling gate -- a {\sc cnot} gate acting on pure input states -- by QPT, maximum-likelihood reconstruction, and analysis of the resulting process matrix. The maximum likelihood technique overcomes the problem that the na\"{\i}ve matrix inversion procedure in QPT, when performed on real (i.e., inherently noisy) experimental data, typically leads to an unphysical process matrix. In a previous maximum-likelihood QPT experiment \cite{mitchell}, a reduced set of fitting constraints was used. Here we present a fully-constrained fitting technique that can be applied to any physical process. After obtaining our physical process matrix, we can accurately determine the action of the gate on any arbitrary input state, including the amount of mixture added and the change in entanglement. We also evaluate useful measures of gate performance. 

The {\sc cnot} gate we characterize, in which two qubits are encoded in the polarization of two single photons, is a nondeterministic gate operating with coincident detection (the gate operation presented here is improved from~\cite{QTnature}). The gate is known to have failed whenever one photon is not detected at each of the two gate outputs, and we postselect against these failure modes. This gate, described in detail in Ref. \cite{QTnature}, produces output states that have high fidelity with the ideal {\sc cnot} outputs, including highly entangled states.

The idea of QPT \cite{chuangQPT,poyatos,mikenike} is to determine a completely positive map $\mathcal{E}$, which represents the process acting on an arbitrary input state $\rho$: 
\begin{equation}
\label{eq:chidef}
\mathcal{E}(\rho)=\sum_{m,n=0}^{d^2-1} \chi_{mn}\hspace{3pt} {\hat A}_m \rho {\hat A}^\dag_n,
\end{equation}
where the ${\hat A_m}$ are a basis for operators acting on $\rho$. The matrix $\chi$ completely and uniquely describes the process $\mathcal{E}$, and can be reconstructed from experimental tomographic measurements. One performs a set of measurements (quantum state tomography \cite{dfvjQST}) on the output of an $n$-qubit quantum gate, for each of a set of inputs. The input states and measurement projectors must each form a basis for the set of $n$-qubit density matrices, requiring $d^2=2^{2n}$ elements in each set \cite{mikenike,qptnote1}. For a two-qubit gate ($d^2=16$), this requires 256 different settings of input states and measurement projectors. An alternative is {\em ancilla assisted process tomography} ~\cite{dariano,altepeter,deMartini},  where $d^2$ separable inputs can be replaced by a suitable single input state from a $d^2$-dimensional Hilbert space. 

Standard QPT reconstruction techniques typically lead to an unphysical process matrix. This is a significant problem, as the predictive power of the process matrix is questionable if it predicts unphysical gate output states. For physicality, it is necessary that the map be {\em completely positive} and not increase the trace. The tomographic data can be used to obtain a {\em physical} process matrix by finding a positive, Hermitian matrix $\tilde{\chi}$ that is the closest fit in a least-squares sense, and subject to a further set of constraints \cite{mikenike} required to make sure that $\tilde{\chi}$ represents a trace-preserving process \cite{qptnote4}: $\sum_{mn}\tilde{\chi}_{mn} {\hat A}^{\dag}_{n} {\hat A}_{m} =I.$ Practically this is achieved by writing a Hermitian parametrization $\vec{t}$ (Ref. \cite{dfvjQST}) of $\tilde{\chi}$, and minimizing the function
\begin{eqnarray}
\label{eq:fitchi}
&&f(\vec{t})= \nonumber\\
&& \sum_{a,b=1}^{d^2}
\frac{1}{\mathcal{C}}\left( c_{ab}-
\mathcal{C}\displaystyle{\sum_{m,n=0}^{d^2-1}}
\langle\psi_{b} |{\hat A}_{m}|\phi_a\rangle
\langle \phi_a|{\hat A}_{n}|\psi_b\rangle
\tilde{\chi}_{mn}(\vec{t})
\right)^2 \nonumber\\
&&+\hspace{3pt}\lambda\left(\sum_{m,n,k=0}^{d^2-1} \tilde{\chi}_{mn}(\vec{t})\hspace{3.5pt} {\rm Tr}\left[{\hat A}_{n} {\hat A}_{k} {\hat A}_{m} \right] - \delta_{k,0}\right),
\end{eqnarray}
where $|\phi_a\rangle$ is the $a^{th}$ input state, $|\psi_b\rangle$ is the $b^{th}$ measurement analyzer setting, $c_{ab}$ is the measured number of coincident counts for the $a^{th}$ input and $b^{th}$ analyzer setting, $\mathcal{C}$ is the total number of coincident photon pairs within the counting time, $\lambda$ is a weighting factor and $\delta$ is the Kronecker delta. The first sum on the right represents a least-squares fit of the Hermitian matrix to the data, and the second enforces the set of further constraints when the $A_{k}$ are elements of the Pauli basis. The parameter $\lambda$ can be adjusted to ensure that the matrix is arbitrarily close to a completely positive map. The technique is architecture independent -- the photon counts can be replaced with the relevant measurement probabilities for any gate realization. Using our {\sc cnot} data, we applied a global numerical minimization technique \cite{qptnote7} to find the minimum of $f(\vec{t})$ (an alternative maximum-likelihood procedure was given in \cite{jezek}).

\begin{figure*}[!htb]
\center{\epsfig{figure=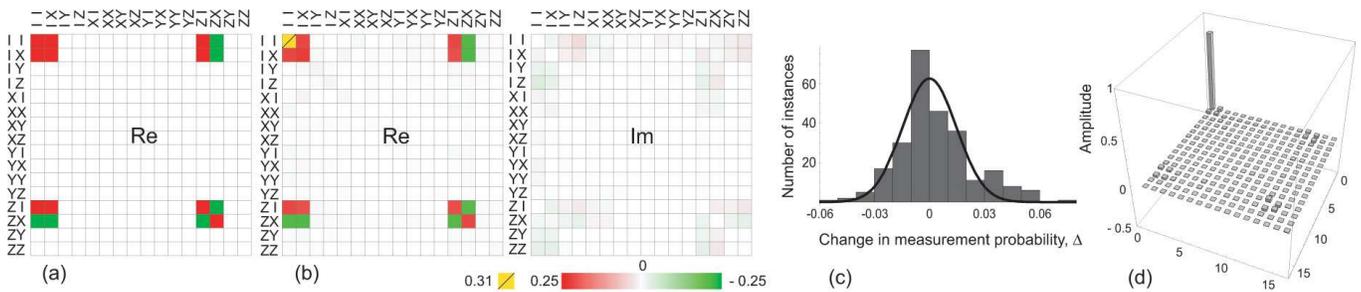,width=510 pt}}
\caption{(color) Process matrix of the {\sc cnot} gate~\cite{epaps}. (a) Ideal process matrix, in the basis defined by tensor products of Pauli operators (imaginary part is identically zero). (b) Maximum-likelihood experimental process matrix. (c) Histogram of the differences in probability between experimental data and the maximum-likelihood reconstruction for the 256 tomographic measurements. The Gaussian fit $\gamma \exp(-\Delta^2/\sigma^2)$ has $\sigma=0.021$. (d) Real part of the experimentally determined process matrix, expressed in the {\sc cnot} operator basis, where the 00 element represents the component of $U_{\rm CNOT}$ in the process. The elements' imaginary parts are negligible, except for one coherence of magnitude 0.06. In this basis, an ideal {\sc cnot} has a unit-valued 00 element; all other elements are zero. The abscissae are the abscissae of (a) and (b), multiplied by $\hat{U}_{\rm CNOT}.$}
\label{fig:chiPauli}
\end{figure*}

The ideal {\sc cnot} can be written as a coherent sum: ${\hat U}_{\rm CNOT}=\tfrac{1}{2}(I \otimes I + I \otimes X + Z \otimes I - Z \otimes X);$ of tensor products of Pauli operators $\{I, X, Y, Z \}$ acting on control and target qubits respectively. The Pauli basis representation of the ideal {\sc cnot}, and our experimental process, are shown in Figs.~\ref{fig:chiPauli}(a) \& \ref{fig:chiPauli}(b). Physically, the process matrix shows the populations of, and coherences between, the basis operators making up the gate function (note the sign of the coherences corresponds to the sign of the terms in $\hat{U}_{\rm CNOT}$), analogous to the interpretation of density matrix elements as populations of, and coherences between, basis quantum states. In fact, process matrices are isomorphic with density matrices in a higher dimensional Hilbert space \cite{jamiolkowski,AGmeasures}, and the trace-preservation condition constrains physical process matrices to a subspace of physical state density matrices.  

How well does the matrix $\tilde{\chi}$ describe the raw data? Clearly there will be some discrepancy, as the simple matrix inversion (i.e. without maximum-likelihood estimation) produces an unphysical process matrix. It is possible to obtain information about the confidence of the fit by examining the {\em residuals} (Fig. \ref{fig:chiPauli}(c)), i.e. the differences $\Delta$ between each of the 256 measurement probabilities and the corresponding probabilities predicted from $\tilde{\chi}$. The width of this distribution, $\sigma(\Delta)=0.021$, gives an idea of the relative error in the process tomography. We further test the maximum-likelihood technique by comparing the predicted output state predicted by $\tilde{\chi}$) with the {\em experimentally determined} output state for all 16 inputs.  The average fidelity and standard deviation between the predicted and measured density matrices are 0.95 and 0.03, respectively. 

Ultimately, we want to characterize the process relative to some ideal: in this case, $\chi_{\rm \textsc{cnot}}$, which is the process matrix representing $\hat{U}_{\rm CNOT}$ \cite{cnotnote}. We use the {\em process fidelity} \cite{AGmeasures}, $F_P={\rm Tr} \left(\chi_{\rm \textsc{cnot}} \hspace{2pt} \tilde{\chi} \right)$, and find $F_P=0.87$. We can obtain a graphical representation of $F_P$ by expressing the process matrix in the {\sc ``cnot''} basis (obtained by acting $\hat{U}_{\rm CNOT}$ on all the Pauli basis elements).  In this case, $F_P$ is simply the height of the corner (00) element, as shown in Fig.~\ref{fig:chiPauli}(d). Currently, we are not able to put an error bar on $F_P$ when it is calculated from $\tilde{\chi}$. The only known technique for obtaining error estimates on the elements of the process matrix comes from performing many reconstructions of the process matrix with random noise added to the raw data in each case \cite{mitchell}. The incorporation of the extra constraints for trace-preservation slows down the numerical minimization to the point where it is impractical to consider repeating the fitting procedure for many simulated data sets. 

The fact that the fidelity of the process is given by the height of one element of $\tilde{\chi}$ in the {\sc cnot} basis suggests that $F_P$ might be obtained with far fewer experimental settings than for full QPT. In principle, only $d^2$ parameters are required to find $F_P$. For our (physically achievable) settings \cite{qptnote1}, the process fidelity with the ideal {\sc cnot} can be calculated directly from a 71-element subset of the tomographic data. Importantly, any such ``direct'' relationships \cite{barbieri} also allow straightforward error estimates. Using this alternative technique, we find  $F^{\prime}_P=0.93 \pm 0.01$ \cite{qptnote6}. The error bar is smaller than $F^{\prime}_P - F_P$, however, the error in $F_P$ is not presently known. 

The {\em average gate fidelity} $\overline{F}$ \cite{MANavgfidel} is defined as the state fidelity \cite{qptnote3} between the actual and ideal gate outputs, averaged over all input states. There is a simple relationship between the process fidelity and the average fidelity for any process \cite{AGmeasures}, which we apply with our data: $\overline{F}^{\prime}=(d \hspace{1pt} F_P^{\prime}+1)/(d+1)=0.95 \pm 0.01.$ We believe that the sub-unit fidelity primarily arises from imperfect mode matching (spatial and spectral overlap of the optical beams). Mode mismatch results in imperfect nonclassical interference between control and target photons, and mixture of the individual qubit states as well. Mode mismatch is not a fundamental limitation for optical quantum gates, and guided mode implementations promise an elegant solution. Nonclassical interferences with $> 99\%$ visibility have been observed with single photons and guided mode beamsplitters \cite{pittmanPRL90}. 

Although the fidelity may seem like a simple method for comparing processes, it is not ideal, because it does not satisfy many of the requirements for a good measure. A full list of the desirable properties can be found elsewhere (e.g.~Ref.~\cite{AGmeasures}), but to some extent they can be summarized by the concept that an ideal measure must remain valid when used to characterize a gate as part of a larger quantum circuit, as well as in isolation. To this end, a more appropriate measure, $C_P=\sqrt{1-F^{\prime}_P}$ has been developed.  Although monotonically related to the process fidelity, it has all the properties required. $C_P$ is a {\em metric}, so that two processes that are identical will have $C_P=0$; orthogonal processes have $C_P=1$. The operational interpretation of $C_P$ is that the average probability of error $\overline{P}_E$ for a quantum computer circuit used to compute some function obeys $\overline{P}_E \leq {C_P}^2$ \cite{AGmeasures}. For our gate, $\overline{P}_E \leq 0.07 \pm 0.01$.

We also introduce a simple relation to characterize how much mixture our gate introduces (for details, see\cite{AGmeasures}): $\overline{{\rm Tr} \left( \rho^2 \right)}=(d \hspace{1pt} {\rm Tr}(\tilde{\chi}^2)+1)/(d+1)=0.83,$ where the quantity on the left hand side is the purity of gate output states, averaged over all pure inputs. This corresponds to an average normalized linear entropy \cite{whiteTangle} of 0.22. 

\begin{figure}[!htb]
\center{\epsfig{figure=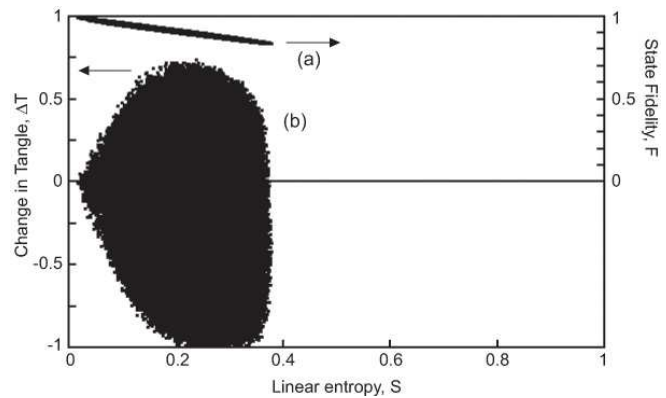,width=\columnwidth}}
\caption{(a) State fidelity of our {\sc cnot} gate outputs (with ideal {\sc cnot} output states) calculated from $\tilde{\chi}$, plotted against the linear entropy added by the gate. A perfect experimental {\sc cnot} process would have F=1, S=0 for all states. (b) Change in tangle between input and output, and linear entropy added, for our {\sc cnot} gate outputs, calculated from $\tilde{\chi}$. An ideal {\sc cnot} would have points with $\Delta T$  distributed between -1 and 1, and S=0.}
\label{fig:scatter}
\end{figure}

An instructive method for examining the action of the gate, in terms of fidelity, entanglement (quantified by the tangle T \cite{wootters,whiteTangle}) and entropy, is to make scatter plots of these quantities for output states of the gate (Fig. \ref{fig:scatter}). We used $\sim$200000 pure, uniformly distributed (by the H\'{a}ar measure) input states, and the $\tilde{\chi}$ matrix, to predict a distribution of output states of our experimental gate. From this data, we can observe the relationship between the mixture added, and both the entanglement generated by the gate and the gate output state fidelity. The gate has three separate spatial mode matching conditions \cite{QTnature}, and the contribution of each of these to the overall mixture is state dependent, leading to the distribution of entropies. There is a clear correlation between the fidelity and the amount of mixture added, and the minimum output state fidelity is 0.83. The shape of the upper lobe ($\Delta T > 0$) of the $\Delta T\hspace{2pt}{\rm vs.}\hspace{2pt}S$ plot (Fig.~\ref{fig:scatter}(b)) is readily understood by the state-dependent mode matching considerations. States that have the largest change in tangle correspond to cases when the gate requires all three mode matching conditions to be simultaneously satisfied, and since each is not perfectly satisfied, this introduces some mixture. There exist partially entangled input states for which $\Delta T \approx 0$ and all three mode matching conditions apply, and these also have higher entropy. When only one mode matching condition applies, the gate cannot perform an entangling operation, but only a little mixture is added. The extension of the lower lobe ($\Delta T <0$) to $\Delta T=-1$ (asymmetric with the upper lobe) can be explained by the fact that when the gate acts to disentangle the input, the addition of mixture also reduces the tangle. We find that the maximum increase in tangle of the gate (the entangling capability \cite{poyatos}) is $\Delta T_{\rm max}=0.73$. 

In summary, we have demonstrated the full characterization of a two-qubit entangling quantum process -- a controlled-{\sc NOT} gate -- by applying physical quantum process tomography. With the process matrix, we can predict, with approximately $95\%$ fidelity, the action of the gate on an arbitrary two-qubit input state. We determine: an average gate fidelity of 0.90 using the process matrix, and $0.95 \pm 0.01$ using a set of 71 input and measurement settings; an average error probability bounded above by 0.07 $\pm$ 0.01; and a maximum increase in tangle of 0.73. The main failure mechanism of the gate can be observed from the process matrix in the Pauli basis, and the scatter plots -- some of the operator population is incoherently redistributed so that the gate performs the identity operation with higher probability than for the ideal {\sc cnot}, a mechanism that we assign primarily to the imperfect mode matching of the interferometers. 

We thank M.J. Bremner, J.S. Lundeen, M.W. Mitchell, M.A. Nielsen, and S. Schneider for helpful discussions. This work was supported in part by the Australian Research Council, and the NSA and ARDA under ARO Contract No. DAAD-19-01-1-0651. AG acknowledges support from the NZ FRST. DFVJ acknowledges support from the MURI Center for Photonic Quantum Information Systems (ARO/ARDA program DAAD19-03-1-0199).


\begin{thebibliography}{34}
\bibitem[{qpt({\natexlab{b}})}]{qptnote4}
\bibinfo{note}{The postselective nature of our experiments precludes maps that
  are not trace-preserving (e.g., where one or more photons are lost). For non-trace-preserving processes, $\sum_{mn}\tilde{\chi}_{mn} {\hat A}^{\dag}_{n} {\hat A}_{m} < I.$}
\bibitem[{qpt({\natexlab{c}})}]{qptnote7}
\bibinfo{note}{We use the NMinimize routine in ${\rm Mathematica}\textsuperscript{\textregistered}\hspace{2pt} 5.0.$}
\end{thebibliography}
\end{document}